\documentclass [12pt] {article}
\usepackage{a4}

  \def\pp{{\mathchoice
          {
              \kern 1pt%
              \raise 1pt
              \vbox{\hrule width5pt height0.4pt depth0pt
                    \kern -2pt
                    \hbox{\kern 2.3pt
                          \vrule width0.4pt height6pt depth0pt
                          }
                    \kern -2pt
                    \hrule width5pt height0.4pt depth0pt}%
                    \kern 1pt
           }
            {
              \kern 1pt%
              \raise 1pt
              \vbox{\hrule width4.3pt height0.4pt depth0pt
                    \kern -1.8pt
                    \hbox{\kern 1.95pt
                          \vrule width0.4pt height5.4pt depth0pt
                          }
                    \kern -1.8pt
                    \hrule width4.3pt height0.4pt depth0pt}%
                    \kern 1pt
            }
            {
              \kern 0.5pt%
              \raise 1pt
              \vbox{\hrule width4.0pt height0.3pt depth0pt
                    \kern -1.9pt  
                    \hbox{\kern 1.85pt
                          \vrule width0.3pt height5.7pt depth0pt
                          }
                    \kern -1.9pt
                    \hrule width4.0pt height0.3pt depth0pt}%
                    \kern 0.5pt
            }
            {
              \kern 0.5pt%
              \raise 1pt
              \vbox{\hrule width3.6pt height0.3pt depth0pt
                    \kern -1.5pt
                    \hbox{\kern 1.65pt
                          \vrule width0.3pt height4.5pt depth0pt
                          }
                    \kern -1.5pt
                    \hrule width3.6pt height0.3pt depth0pt}%
                    \kern 0.5pt
            }
        }}

  \def\mm{{\mathchoice
                       {
                             \kern 1pt
               \raise 1pt    \vbox{\hrule width5pt height0.4pt depth0pt
                                  \kern 2pt
                                  \hrule width5pt height0.4pt depth0pt}
                             \kern 1pt}
                       {
                            \kern 1pt
               \raise 1pt \vbox{\hrule width4.3pt height0.4pt depth0pt
                                  \kern 1.8pt
                                  \hrule width4.3pt height0.4pt depth0pt}
                             \kern 1pt}
                       {
                            \kern 0.5pt
               \raise 1pt
                            \vbox{\hrule width4.0pt height0.3pt depth0pt
                                  \kern 1.9pt
                                  \hrule width4.0pt height0.3pt depth0pt}
                            \kern 1pt}
                       {
                           \kern 0.5pt
             \raise 1pt  \vbox{\hrule width3.6pt height0.3pt depth0pt
                                  \kern 1.5pt
                                  \hrule width3.6pt height0.3pt depth0pt}
                           \kern 0.5pt}
                       }}

\catcode`@=11
\def\un#1{\relax\ifmmode\@@underline#1\else
        $\@@underline{\hbox{#1}}$\relax\fi}
\catcode`@=12

\let\du=\du                     

\def\a{\alpha}
\def\b{\beta}

\def\d{\delta}

\def\f{\phi}
\def\g{\gamma}
\def\h{\eta}

\def\j{\psi}
\def\k{\kappa}
\def\l{\lambda}
\def\m{\mu}
\def\n{\nu}

\def\p{\pi}
\def\q{\theta}
\def\r{\rho}
\def\s{\sigma}
\def\t{\tau}

\def\z{\zeta}
\def\D{\Delta}

\def\G{\Gamma}

\def\L{\Lambda}
\def\O{\Omega}

\def\Q{\Theta}

\def\ve{\varepsilon}
\def\vf{\varphi}

\def\ch{{\cal H}}

\def\cm{{\cal M}}

\def\bo{{\raise-.5ex\hbox{\large$\Box$}}}               
\def\pa{\partial}                                       
\def\de{\nabla}                                         
\def\TH{{\raise.2ex\hbox{$\displaystyle \bigodot$}\mskip-4.7mu \llap H \;}}
\def\face{{\raise.2ex\hbox{$\displaystyle \bigodot$}\mskip-2.2mu \llap {$\ddot
        \smile$}}}                                      

\def\Tilde#1{\widetilde{#1}}                    
\def\Bar#1{\overline{#1}}                       
\def\VEV#1{\left\langle #1\right\rangle}        
\def\abs#1{\left| #1\right|}                    
\def\leftrightarrowfill{$\mathsurround=0pt \mathord\leftarrow \mkern-6mu
        \cleaders\hbox{$\mkern-2mu \mathord- \mkern-2mu$}\hfill
        \mkern-6mu \mathord\rightarrow$}
\def\dvec#1{\vbox{\ialign{##\crcr
        \leftrightarrowfill\crcr\noalign{\kern-1pt\nointerlineskip}
        $\hfil\displaystyle{#1}\hfil$\crcr}}}           

\def\frac#1#2{{\textstyle{#1\over\vphantom2\smash{\raise.20ex
        \hbox{$\scriptstyle{#2}$}}}}}                   
\def\sfrac#1#2{{\vphantom1\smash{\lower.5ex\hbox{\small$#1$}}\over
        \vphantom1\smash{\raise.4ex\hbox{\small$#2$}}}} 
\def\bfrac#1#2{{\vphantom1\smash{\lower.5ex\hbox{$#1$}}\over
        \vphantom1\smash{\raise.3ex\hbox{$#2$}}}}       
\def\afrac#1#2{{\vphantom1\smash{\lower.5ex\hbox{$#1$}}\over#2}}    

\def\[{\lfloor{\hskip 0.35pt}\!\!\!\lceil}
\def\]{\rfloor{\hskip 0.35pt}\!\!\!\rceil}
\def\Lag{{\cal L}}
\def\du#1#2{_{#1}{}^{#2}}

\def\fracm#1#2{\hbox{\large{${\frac{{#1}}{{#2}}}$}}}

\def\un{\underline}
\def\fracmm#1#2{{{#1}\over{#2}}}

\def\low#1{{\raise -3pt\hbox{${\hskip 0.75pt}\!_{#1}$}}}

\def\Tilde#1{{\widetilde{#1}}\hskip 0.015in}

\newskip\humongous \humongous=0pt plus 1000pt minus 1000pt
\def\caja{\mathsurround=0pt}
\def\eqalign#1{\,\vcenter{\openup2\jot \caja
        \ialign{\strut \hfil$\displaystyle{##}$&$
        \displaystyle{{}##}$\hfil\crcr#1\crcr}}\,}
\newif\ifdtup

\def\pl#1#2#3{Phys.~Lett.~{\bf {#1}B} (19{#2}) #3}
\def\np#1#2#3{Nucl.~Phys.~{\bf B{#1}} (19{#2}) #3}

\def\cqg#1#2#3{Class.~and Quantum Grav.~{\bf {#1}} (19{#2}) #3}
\def\cmp#1#2#3{Commun.~Math.~Phys.~{\bf {#1}} (19{#2}) #3}

\parskip=\medskipamount                 
\thicklines                         

\begin{document}
\thispagestyle{empty}

{\hbox to\hsize{
\vbox{\noindent October 2002         \hfill hep-th/0209003  \\
                revised version       }}}

\noindent
\vskip1.3cm
\begin{center}

{\Large\bf Summing up D-Instantons in N=2 Supergravity}

\vglue.2in

                  Sergei V. Ketov

{\it          Department of Physics\\
           Tokyo Metropolitan University\\
          Hachioji, Tokyo 192--0397, Japan}
\vglue.1in
{\sl ketov@phys.metro-u.ac.jp}

\end{center}
\vglue.2in
\begin{center}
{\Large\bf Abstract}
\end{center}

The non-perturbative quantum geometry of the Universal Hypermultiplet (UH) is 
investigated in N=2 supergravity. The UH low-energy effective action is given 
by the four-dimensional quaternionic non-linear sigma-model having an 
$U(1)\times U(1)$ isometry. The UH metric is governed by the single real 
pre-potential that is an eigenfunction of the Laplacian in the hyperbolic 
plane. We calculate the classical pre-potential corresponding to the standard 
(Ferrara-Sabharwal) metric of the UH arising in the Calabi-Yau 
compactification of type-II superstrings. The non-perturbative quaternionic 
metric, describing the D-instanton contributions to the UH geometry, is found 
by requiring the $SL(2,{\bf Z})$ modular invariance of the UH pre-potential. 
The pre-potential found is unique, while it coincides with the D-instanton 
function of Green and Gutperle, given by the order-$3/2$ Eisenstein series. 
As a by-product, we prove cluster decomposition of D-instantons in curved 
spacetime. The non-perturbative UH pre-potential interpolates between the 
perturbative (large CY volume) region and the superconformal (Landau-Ginzburg)
region in the UH moduli space. We also calculate a non-perturbative scalar 
potential in the hyper-K\"ahler limit, when an abelian isometry of the UH 
metric is gauged in the presence of D-instantons. 

\newpage

\section{Introduction}

Instanton corrections in compactified M-theory/superstrings are crucial for 
solving the fundamental problems of vacuum degeneracy and supersymmetry 
breaking. Some instanton corrections in the type-IIA superstring theory 
compactified on a {\it Calabi-Yau} (CY) threefold arise due to the Euclidean 
D2-branes wrapped about the CY special (supersymmetric) three-cycles 
\cite{bbs,bb2}. Being BPS solutions to the Euclidean ten-dimensional (10d) 
supergravity equations of motion, these wrapped branes are localized in four 
uncompactified spacetime dimensions and thus can be identified with instantons.
 They are called D-instantons. The D-instanton action is essentially given by 
the volume of the supersymmetric 3-cycle on which a D2-brane is wrapped.  The 
supersymmetric cycles (by definition) minimize volume in their homology class. 

At the level of the {\it Low-Energy Effective Action} (LEEA), the effective 
field theory is given by the four-dimendsional (4d), N=2 supergravity with 
some N=2 vector- and hyper-multiplets, whose structure is dictated by the CY 
cohomology, and whose moduli spaces are independent. The hypermultiplet sector
of the LEEA is described by a 4d, N=2 {\it Non-Linear Sigma-Model} (NLSM) 
with a quaternionic metric in the NLSM target (moduli)  space \cite{bw}. Any 
CY compactification gives rise to the so-called {\it Universal Hypermultiplet}
 (UH)  in 4d, which contains a dilaton amongst its field components. 

When the type-IIA supergravity 3-form has a non-vanishing CY-valued 
expectation value, the UH becomes electrically charged. This implies that 
an abelian isometry of the NLSM target (= UH moduli) space is gauged, while 
the UH scalar potential is non-trivial \cite{ps,pot}.

It is of considerable interest to calculate the hypermultiplet 
{\it non}-perturbative NLSM metric and the associated scalar potential, 
by including the D-instanton corrections. The qualitative analysis was 
initiated by Witten \cite{w1} who showed that the D-instanton quantum 
corrections are given by powers of $e^{-1/g_{\rm string}}$, 
where $g_{\rm string}$ is the type-II superstring coupling constant \cite{w1}.
The D-instanton actions were computed in refs.~\cite{bb2,gusp,tvv}. The 
$SL(2,{\bf Z})$ modular-invariant completion of the $R^4$-terms by the 
D-instanton effects in the ten-dimensional type-II superstrings was found in
ref.~\cite{gg}. Some of these $R^4$-terms arise from a one-loop calculation in
eleven-dimensional M-theory \cite{gg2}. The D-brane contributions to the 
$R^4$-couplings in any toroidal compactification of type-II superstrings, as 
well as their relation to the Eisenstein series in lower $(7,8,9)$ spacetime 
dimensions, were investigated in ref.~\cite{pki}. The CY wrapped D-branes 
from the mathematical viewpoint are reviewed in ref.~\cite{doug}.

An exact derivation of the non-perturbative hypermultiplet LEEA in a generic CY
superstring compactification is hardly possible, and this is far beyond the 
scope of this paper. Our discussion and the results are limited to the 
{\it Universal Hypermultiplet} (UH) sector present in {\it any} CY superstring 
compactification. The constraints imposed by {\it local} supersymmetry with 
eight conserved supercharges (e.g., N=2 in 4d) on the UH moduli space and  
its metric take the form of quaternionic geometry. In the  case of a {\it 
single} hypermultiplet the quaternionic constraints are given by the 
Einstein-Weyl integrable system of non-linear partial differential equations. 
A generic CY compactification (or the `realistic' superstring models) have 
{\it several} matter hypermultiplets, while the UH and CY moduli are mixed 
and the quaternionic constraints are less restrictive. Our results are
immediately applicable to the {\it `rigid'} CY manifolds characterized by the 
vanishing Hodge number $h_{2,1}=0$. Some explicit realizations of the rigid CY 
manifolds, including the famous ${\cal Z}$ manifold, were given e.g., 
in ref.~\cite{cdp}.  

Having confined ourselves to the UH sector, we still have to deal with the 
very difficult problem of solving the Einstein-Weyl equations. To simplify 
this problem, we restricted ourselves to a calculation of the D-instanton
corrections only. In a generic case, the NS five-branes wrapped about the 
entire CY manifold also contribute to the UH quantum geometry --- see e.g., 
refs.~\cite{bbs,nbi} for details. So our results apply when the five-brane
instantons are absent or suppressed in the UH sector. In particular, the
D-instantons are expected to dominate over the 5-brane instantons when the
string coupling $g_{\it string}$ is sufficiently small \cite{w1}. 

The D-instanton contributions to the metric of a single matter hypermultiplet 
were found by Ooguri and Vafa \cite{ov} in the hyper-K\"ahler limit when both
N=2 supergravity and UH are switched off. The results of ref.~\cite{ov} give 
valuable insights into the structure of D-instanton corrections to the UH. The
gravitational corrections are expected to be important at strong coupling too,
 while the UH sector is a good place for studying them. In particular,
 Strominger \cite{one} proved the absence of {\it perturbative} 
superstring corrections to the local UH metric provided that the Peccei-Quinn 
type isometries of the classical UH metric described by the symmetric space 
$SU(2,1)/SU(2)\times U(1)$ \cite{fsh} are preserved. In our earlier papers 
\cite{plb,pro}  we proposed the procedure for a derivation of the 
non-perturbative UH metric in a curved spacetime. However, no explicit 
quaternionic solution in the expected form of the infinite D-instanton sum 
was given in refs.~\cite{plb,pro}. In this paper we propose such a solution 
by requiring the $SL(2,{\bf Z})$ duality invariance of the Calderbank-Petersen
\cite{cp} pre-potential describing the $U(1)\times U(1)$-symmetric 
four-dimensional quaternionic metrics.

We also turn to the gauged version of the universal hypermultiplet NLSM, 
by gauging  one of its abelian isometries preserved by D-instantons. This 
gives rise to the non-perturbative scalar potential whose minima determine the
`true' vacua in our toy model comprising the UH coupled to the single N=2 
vector multiplet gauging the UH abelian isometry. As is well-known (see, e.g.,
 ref.~\cite{pot}), gauging the classical UH geometry  gives rise to the 
dilaton potential whose minima occur outside of the region where the string 
perturbation theory applies.  However, this potential with the run-away 
behaviour  is not protected against instanton corrections, while it is 
reasonable to gauge only those NLSM isometries that are not broken after 
the D-instanton corrections are included. Because of the brane charge 
quantization, the classical symmetries of the UH metric are generically broken
 by the wrapped D2-branes and the solitionc five-branes wrapped about the 
entire CY \cite{bbs}. However, when merely the D-instantons are taken into 
account, a continuous abelian symmetry of the UH moduli space may survive, 
while it also makes actual calculations possible \cite{nbi}. 

Our paper is organized as follows: in sect.~2 we recall a few basic facts 
about the type-II string dilaton and the 4d NLSM it belongs to. The explicit
relations between various parametrizations of the classical UH metric, used in
the literature, are also presented in sect.~2. In sect.~3 we discuss all 
possible deformations of this metric under the condition of unbroken 4d, N=2 
local supersymmetry, and their relation to integrable models. The {\it 
Calderbank-Pedersen} (CP) linearization procedure for the $U(1)\times U(1)$ 
symmetric quaternionic metrics is also formulated in sect.~3. The Ooguri-Vafa
solution is reviewed in sect.~4. The most general modular-invariant 
quaternionic UH metric, describing the infinite D-instanton sum in N=2 
supergravity, is proposed in sect.~5. Sect.~6 is devoted to the gauged version
 of the UH and its scalar potential in the presence of the D-instanton 
corrected UH metric.
\vglue.2in

\section{Dilaton and NLSM}

In all four-dimensional superstring theories a dilaton scalar $\vf$ is 
accompanied by an axion pseudo-scalar $R$~  
belonging to the same scalar supermultiplet. In the (classical) supergravity 
approximation, their LEEA (or kinetic terms)
 are given by the NLSM whose structure is entirely fixed by duality: 
the NLSM target space is given by the two-dimensional non-compact homogeneous 
 space $SL(2,{\bf R})/U(1)$. In the full `superstring theory' (including 
branes) the continuous symmetry $SL(2,{\bf R})\cong SO(2,1)
\cong SU(1,1)$ is generically broken to a discrete subgroup of $SL(2,{\bf Z})$
 \cite{ht}, whereas the local NLSM metric may receive some non-perturbative 
(instanton) corrections \cite{bbs,w1,nbi,ov,one}. 

The $SL(2,{\bf R})/U(1)$-based NLSM can be parametrized in terms of a single 
complex scalar,
$$ A \equiv A_1+iA_2  = R + ie^{-\vf}~,\eqno(2.1)$$
subject to the $SL(2,{\bf R})$ duality transformations
$$ A~\to~A' =\fracmm{aA +b}{cA +d}~,\quad {\rm where}\quad
\left(\begin{array}{cc} a & b \\ c & d \end{array}\right)\in SL(2,{\bf R})~,
\eqno(2.2)$$
with four real parameters $(a,b,c,d)$ obeying the 
 condition $ad-bc=1$. The $SL(2,{\bf R})$ NLSM Lagrangian in the
parametrization (2.1) is given by
$$ \k^2 \Lag(A,\bar{A}) = \fracmm{1}{(A-\bar{A})^2}\pa^{\m}\bar{A}\pa_{\m}A~.
\eqno(2.3)$$
We assume 
 that our scalars are dimensionless. The dimensional 
coupling constant $\k$ of the UH NLSM is proportional to  
the gravitational coupling constant. We assume that $\k^2=1$ 
for notational simplicity in what follows.

It is easy to check that the NLSM metric defined by eq.~(2.3) is K\"ahler, 
with a K\"ahler potential
$$ K(A,\bar{A}) = \log(A-\bar{A})~.\eqno(2.4)$$
The $SL(2,{\bf R)}$ transformations (2.2) are generated by constant shifts of 
the axion (T-duality) with
$$ \left(\begin{array}{cc} 1 & 1 \\ 0 & 1 
\end{array}\right)\in SL(2,{\bf R})~,\eqno(2.5)$$
and the S-duality transformation $\left(e^{-\vf}\to e^{+\vf}\right)$ with
$$ \left(\begin{array}{cc} 0 & 1 \\ -1 & 0 \end{array}\right)\in 
SL(2,{\bf R})~.\eqno(2.6)$$

It is worth mentioning that the K\"ahler potential is 
 defined modulo K\"ahler gauge transformations,
$$ K(A,\bar{A}) ~\to~  K(A,\bar{A}) +f(A) +\bar{f}(\bar{A})~,\eqno(2.7)$$
with arbitrary (locally holomorphic) functions $f(A)$. After 
 the field redefinition 
$$ S =i\bar{A}\equiv  e^{-2\f}+i2D~, \eqno(2.8)$$
in terms of a dilaton $\f$ and an axion $D$, 
the  K\"ahler potential (2.4) takes the form
$$ K(S,\bar{S})=  \log(S+\bar{S})~.\eqno(2.9)$$
This parametrization was used, for example, in refs.~\cite{bbs,nbi,one}. 

To connect the K\"ahler potential (2.9) to the standard (Fubuni-Study) 
potential used in the mathematical 
literature, let's make yet another field redefinition,
$$ S=\fracmm{1-z}{1+z}~.\eqno(2.10)$$
Then the new K\"ahler potential $K(z,\bar{z})$ takes the dual Fubini-Study form
$$ K(z,\bar{z}) = \log(1-\abs{z}^2)~~.\eqno(2.11)$$
The corresponding NLSM Lagrangian is 
$$ -\Lag(\f,D)= (\pa_{\m}\f)^2 +e^{4\f}(\pa_{\m}D)^2~,\eqno(2.12)$$
or   
$$ -4\Lag(\r,t)= \fracmm{1}{\r^2}\left[ (\pa_{\m}\r)^2 
+(\pa_{\m}t)^2\right]~,\eqno(2.13)$$
where we have introduced the new variables
$$ \r=e^{-2\f} \quad{\rm and}\quad t=2D~.\eqno(2.14)$$
The metric of the NLSM (2.13) is conformally flat, it has a negative scalar 
curvature and the manifest isometry because of the $t$-independence of its 
components.  

The (complex) one-dimensional K\"ahler potential (2.11) has a natural 
(K\"ahler and dual Fubini-Study) extension to two complex dimensions,
$$  K(z_1,z_2,\bar{z}_1,\bar{z}_2)=
\log(1-\abs{z_1}^2 -\abs{z_2}^2)~~,\eqno(2.15)$$
where $(z_1,z_2)\in {\bf C}^2$ are on equal footing inside the ball $B^4$:  
 $\abs{z_1}^2 +\abs{z_2}^2 < 1$.  
The K\"ahler potential (2.15) defines the so-called Bergmann metric 
$$ ds^2= \fracmm{dz_1d\bar{z}_1+dz_2d\bar{z}_2}{1-\abs{z_1}{}^2-\abs{z_2}{}^2}
+ \fracmm{(\bar{z}_1dz_1+\bar{z}_2dz_2)(z_1d\bar{z}_1+z_2d\bar{z}_2)}{(1-
\abs{z_1}{}^2-\abs{z_2}{}^2)^2} \eqno(2.16)$$
in the open ball $B^4$. Being equipped with the Bergmann metric, the open ball
 $B^4$ is equivalent to the (non-compact) symmetric quaternionic space 
$SU(2,1)/U(2)$ \cite{besse} that is dual to the compact projective space 
${\bf CP}^2=SU(3)/SU(2)\times U(1)$ \cite{cp2}. The homogeneous space 
${\bf CP}^2$ is symmetric, while it is also an Einstein space of positive 
scalar curvature with the (anti)self-dual Weyl tensor.~\footnote{The Weyl 
tensor is the traceless part of the Riemann curvature.} The non-compact coset 
space $SU(2,1)/U(2)$ is, therefore, also an Einstein space (though of negative
 scalar curvature), with the (anti)self-dual Weyl tensor. In other words, the 
coset space $SU(2,1)/U(2)$ is an Einstein-Weyl (or a self-dual Einstein) 
space. The four-dimensional Einstein-Weyl spaces are called {\it quaternionic}
 by definition \cite{besse}. 

The relation between the Bergmann parametrization (2.16), used in the 
mathematical literature, and the UH parametrization 
$(\f,D,C,\bar{C})$, used in the physical literature \cite{bbs,fsh,nbi}, 
is given by
$$ z_1= \fracmm{1-S}{1+S}~,\quad  z_2= \fracmm{2C}{1+S}~~~,\eqno(2.17)$$
where the new complex variable $C$ can be identified with a RR-scalar of UH, 
whereas another complex scalar $S$ of UH is now given by  
({\it cf.} eq.~(2.8))
$$ S= e^{-2\f}+i2D +\bar{C}C~.\eqno(2.18)$$
Two complex scalars $(S,C)$ represent all bosonic physical degrees of freedom
in UH that also has a Dirac hyperino as the fermionic superpartner. 
The K\"ahler
 potential of the UH metric in terms of the new coordinates $(S,C)$ reads
$$ K(S,\bar{S}, C,\bar{C}) = \log\left(S+\bar{S}-2C\bar{C}\right)~,
\eqno(2.19)$$
while the corresponding metric is given by
$$ ds^2= e^{2K}\left(dSd\bar{S} -2CdSd\bar{C}-2\bar{C}d\bar{S}dC+
2(S+\bar{S})dCd\bar{C}\right)~.\eqno(2.20)$$

The bosonic part of the NLSM Lagrangian of UH in terms of the 
scalar fields  $(\f,D,C,\bar{C})$ reads 
$$ - \Lag_{\rm FS} = (\pa_{\m}\f)^2 + e^{2\f}\abs{\pa_{\m}C}^2 
+ e^{4\f}(\pa_{\m}D +\fracm{i}{2}\bar{C}
\dvec{\pa_{\m}}C)^2~.~\eqno(2.21)$$
The metric of this NLSM is diffeomorphism-equivalent to the quaternionic 
Bergmann metric on $SU(2,1)/U(2)$ by our construction. At the same time, 
eq.~(2.21) coincides with the so-called {\it Ferrara-Sabharwal} (FS) NLSM 
(in the physical literature) that was derived \cite{fsh} by compactifying the 
10d type-IIA supergravity on a CY threefold in the universal (UH) sector down 
to four spacetime timensions, $\m=0,1,2,3$. This means that we can 
identify our field $\f$ with the dilaton used in refs.~\cite{bbs,fsh}. The FS 
metric in eq.~(2.21) is thus completely fixed by the duality symmetries of 
$SU(2,1)/SU(2)\times U(1)$ \cite{nbi}. The FS metric can be trusted
as long as the string coupling is not strong, 
$g_{\rm string}=e^{\VEV{\f}}=\VEV{1/\sqrt{\r}}$, i.e. for large $\r>0$.  
The variable $\r$ has the physical meaning of the CY space volume
 --- see eq.~(2.14) and ref.~\cite{one}.

After the coordinate change
$$ z_1=r\cos\fracm{\q}{2}e^{i(\vf+\j)/2}~,\quad
 z_2=r\sin\fracm{\q}{2}e^{-i(\vf-\j)/2}~,\eqno(2.22)$$
the Bergmann metric (2.16) can be rewritten to the diagonal form in the 
Bianchi IX formalism with manifest $SU(2)$ symmetry,
$$ ds^2=\fracmm{dr^2}{(1-r^2)^2} +\fracmm{r^2\s^2_2}{(1-r^2)^2}+
\fracmm{r^2}{(1-r^2)}(\s_1^2+\s_3^2)~,\eqno(2.23)$$
where we have introduced the $su(2)$ (left)-invariant one-forms \cite{nbi}
$$ \eqalign{
\s_1= & -\fracm{1}{2} \left( \sin\j\sin\q d\vf +\cos\j d\q\right)~,\cr
\s_2= & \fracm{1}{2} \left( d\j +\cos\q d\vf\right)\,\cr
\s_3= & \fracm{1}{2} \left( \sin\j d\q  -\cos\j\sin\q d\vf\right)~,\cr}
\eqno(2.24)$$
in terms of four real coordinates $0\leq r<1$,  $0\leq \q < \p$,  
$0\leq \vf< 2\p$ and $0\leq \j< 4\p$. The one-forms (2.24) obey the relations
$$ \s_i\wedge \s_j =\fracm{1}{2}\ve_{ijk}d\s_k~,\quad i,j,k=1,2,3~.
\eqno(2.25)$$

Yet another parametrization of the same metric arises after the coordinate 
change
$$ r= \sqrt{ \fracmm{2}{\tilde{s}+1} }~~~~,\eqno(2.26)$$
where $1< \tilde{s} < \infty$. This yields 
$$ ds^2= \fracmm{1}{(\tilde{s}-1)^2}\left[
\fracmm{1}{4(\tilde{s}+1)}d\tilde{s}^2 + 2(\tilde{s}-1)(\s^2_1 +\s^2_3) +
2(\tilde{s}+1)\s^2_2\right]~.\eqno(2.27)$$
This metric appeared in the literature as the special case of the 
so-called {\it quaternionic Eguchi-Hanson} family of four-dimensional 
quaternionic metrics with a $U(1)\times U(1)$ isometry, being derived from 
harmonic superspace by Ivanov and Valent \cite{iv,iva} --- 
see eqs.~(3.49) and (4.7) of ref.~\cite{iv} and subsect.~5.5 of 
ref.~\cite{iva}.~\footnote{In our notation versus the one of 
refs.~\cite{iv,iva}, $\tilde{s}$ has to be rescaled by a gravitational 
coupling \newline ${~~~~~}$ constant $\k^2$, the one-forms 
$\s_i$ are to be rescaled by a factor of $2\sqrt{2}$, while $\s_2$ and 
$\s_3$ are to be \newline ${~~~~~}$ interchanged.} 

One of the technical problems in dealing with hypermultiplets and the 
quaternionic geometry (vs. N=2 vector multiplets and the special geometry)
is the need of complicated (non-holomorphic) field reparametrizations when 
going from one coordinate frame to another. For  example, it is common in the 
physical literature to use complex coordinates $(C,S)$, as in eq.~(2.19), in 
order to make manifest the K\"ahler nature of the classical UH metric. As will
be shown in the next sections, the D-instanton corrections in {\it curved} 
spacetime do {\it not} preserve the K\"ahler nature of the hypermultiplet 
metric, though they do preserve its quaternionic nature. Hence, complex 
coordinates are not useful for D-instantons. 

\section{D-instantons and quaternionic geometry}  

Quantum non-perturbative corrections generically break all the continuous 
$SU(2,1)$ symmetries of the UH classical NLSM down to a discrete subgroup 
because of  charge quantization, even if local N=2 supersymmetry 
in 4d remains unbroken \cite{bbs}. However, in some special non-trivial 
situations, part of the continuous abelian symmetries of the UH moduli space 
may survive. As was demonstrated by Strominger \cite{one}, there
 is no non-trivial quaternionic deformation of the classical FS metric in the 
superstring perturbation theory when {\it all} the Peccei-Quinn-type 
symmetries (with three real parameters $(\a,\b,\g)$), 
$$ D\to D+\a~,\quad C\to C+ \g -i\b~,\quad S\to S +2(\g+i\b)C+\g^2+\b^2~,
\eqno(3.1)$$
remain unbroken. However, {\it some} of the Peccei-Quinn-type symmetries 
(namely, the one associated with constant shifts of the RR scalar $C$) can be 
broken non-perturbatively (namely, by D-instantons) \cite{nbi}. In this 
paper we assume that the abelian isometry, associated with constant shifts of 
the axionic $D$-field, is preserved since we only consider D-instantons and
ignore five-brane instantons. We also assume, like in refs.~\cite{ov,plb}, 
that the abelian rotations of the $C$ field are preserved by D-instantons
too (see sect.~4). The surviving symmetries are
$$ D\to D +\a \quad{\rm and}\quad  C\to e^{2\p i\d}C~,\quad{\rm where}\quad 
 \d\in [0,1]~~.
 \eqno(3.2)$$ 
Our assumption of the $U(1)\times U(1)$ isometry of the non-perturbative UH 
moduli space metric is consistent with the known $U(1)\times U(1)$ symmetry of
the Ooguri-Vafa solution (see sect.~4) describing the D-instanton corrections 
to a matter hypermultiplet \cite{ov}. Further support for this symmetry comes
from the type-IIB side, where the N=2 {\it double-tensor} multiplet version of
UH has to be used in the Euclidean path-integral approach to D-instantons 
\cite{tvv}. The Euclidean N=2 double-tensor multiplet action is bounded
from below, while its  $U(1)\times U(1)$ symmetry is obviously protected 
against quantum corrections \cite{tvv}.

Our considerations are entirely local, while the D-instanton corrected UH 
metric is supposed to be quaternionic (as long as local N=2 supersymmetry is 
preserved) with an $U(1)$ or an $U_D(1)\times U_C(1)$ (torus) isometry. Our 
problem now amounts to a derivation of non-trivial quaternionic deformations 
of the Bergmann (or FS) metric, which can be physically interpreted as the 
D-instantons coming from the wrapped D2-branes, subject to a given abelian 
isometry.

A generic quaternionic manifold admits three independent {\it almost} complex 
structures $(\tilde{J}_A)\du{a}{b}$, where $A=1,2,3$ and $a,b=1,2,3,4$. Unlike
the hyper-K\"ahler manifolds, the quaternionic complex structures are 
{\it not} covariantly constant, i.e. they are not integrable to some complex 
structures because of a non-vanishing torsion. This NLSM torsion is induced by
4d, N=2 supergravity because the quaternionic condition on the hypermultiplet 
NLSM target space metric is the direct consequence of local N=2 supersymmetry 
in four spacetime dimensions \cite{bw}. As regards real {\it four}-dimensional
 quaternionic manifolds (relevant to the UH), they all have 
{\it Einstein-Weyl\/} geometry of {\it negative\/} scalar curvature 
\cite{bw,besse}, 
$$ W^-_{abcd}=0~,\qquad R_{ab}=-\fracmm{\L}{2}g_{ab}~,\eqno(3.3)$$  
where $W_{abcd}$ is the Weyl tensor, $R_{ab}$ is the Ricci tensor of the 
metric $g_{ab}$, and the constant $\L>0$ is 
proportional to the gravitational coupling constant. The precise value of the
 `cosmological constant' $\L$ in our notation is fixed in eq.~(3.13). 

Since we assume that the UH quaternionic metric has at least one abelian 
isometry, a good starting point is the Tod theorem
\cite{tod} applicable to {\it any} Einstein-Weyl metric of a non-vanishing 
scalar curvature with a Killing vector $\pa_t$. 
According to ref.~\cite{tod}, there exists a parametrization $(t,\r,\m,\n)$ in
 which such metric takes the form
$$ ds^2_{\rm Tod}= \fracmm{1}{\r^2}\left\{ \fracmm{1}{P}(dt+\hat{\Q})^2 
+P\left[ e^u(d\m^2+d\n^2)+d\r^2\right]\right\}~~,\eqno(3.4)$$
in terms of two potentials, $P$ and $u$, and the one-form $\hat{\Q}$ 
in three dimensions $(\r,\m,\n)$. The first 
potential $P(\r,\m,\n)$ is fixed in terms of the second potential $u$ as 
\cite{tod} 
$$ P= \fracmm{3}{2\L} \left(\r\pa_{\r}u-2\right)~~.\eqno(3.5a)$$
The potential $u(\r,\m,\n)$ itself obeys the 3d {\it non-linear} equation 
$$ (\pa^2_{\m}+\pa^2_{\n})u+\pa^2_{\r}(e^u)=0 \eqno(3.5b)$$
known as the (integrable) $SU(\infty)$ or 3d Toda equation, whereas 
the one-form $\hat{\Q}$ satisfies the {\it linear} differential 
equation 
$$-d\wedge\hat{\Q} =(\pa_{\n}P) d\m\wedge d\r +(\pa_{\m}P) d\r\wedge d\n 
+\pa_{\r}(Pe^u) d\n\wedge d\m~.\eqno(3.5c)$$   

Some comments are in order. 

Given an isometry of the quaternionic metric $g_{ab}$ 
with a Killing vector $K^{a}$, 
$$ K^{a;b} + K^{b;a} =0~, \quad K^2\equiv g_{ab}K^aK^b \neq 0~,\eqno(3.6)$$
we can always choose some adapted coordinates, with  all the metric components
 being  independent upon one of the coordinates $(t)$, as in eq.~(3.4). We can
 then  plug the Tod Ansatz (3.4) into the Einstein-Weyl equations (3.3). It 
follows that this precisely amounts to the equations (3.5). We verified this 
claim \cite{tod} by the use of Mathematica.

It is worth mentioning that after the conformal rescaling 
$$ g_{ab}\to \r^2 g_{ab}\eqno(3.7)$$
a generic Einstein-Weyl metric of the form (3.4) becomes K\"ahler with the 
vanishing scalar curvature \cite{gau}. After this 
conformal rescaling the Tod Ansatz (3.4) precisely takes the form of the 
standard (LeBrun) Ansatz for scalar-flat K\"ahler 
metrics \cite{lebrun}, 
$$ ds^2_{\rm LeBrun}= \fracmm{1}{P}(dt+\hat{\Q})^2 
+P\left[ e^u(d\m^2+d\n^2)+d\r^2\right]~~,\eqno(3.8)$$
whose potential $u$ still satisfies the 3d Toda equation (3.5b), whereas the 
potential $P$ is a solution to 
$$ (\pa^2_{\m} + \pa^2_{\n})P + \pa^2_{\r}(e^uP)=0~.\eqno(3.9)$$
This equation is nothing but the integrability condition for eq.~(3.5c) 
that holds too.

According to LeBrun \cite{lebrun}, a scalar-flat K\"ahler metric is 
{\it hyper}-K\"ahler if and only if
$$ P \propto \pa_{\r}u~.\eqno(3.10)$$
Given eq.~(3.10), the LeBrun Ansatz reduces to the Boyer-Finley Ansatz 
\cite{bf} for a four-dimensional hyper-K\"ahler metric 
with a rotational isometry \cite{bf}, or to the Gibbons-Hawking Ansatz 
\cite{gh} in the case of a translational (or tri-holomorphic) isometry 
that essentially implies $u=0$ in addition. Both Ans\"atze are well known 
in general relativity (see, e.g., ref.~\cite{book} for
a review). In particular, exact solutions to the Boyer-Finley Ansatz are 
 governed by the same 3d Toda (non-linear) equation, whereas
exact solutions to the Gibbons-Hawking Ansatz \cite{gh}
$$ ds^2_{\rm GH}= \fracmm{1}{P}(dt+\hat{\Q})^2 +P(d\m^2+d\n^2+d\r^2) 
\eqno(3.11)$$
are governed by the {\it linear} equations, $(\pa^2_{\m} + \pa^2_{\n} 
+\pa^2_{\r})P=0$ and $\vec{\de}P+\vec{\de}\times\vec{\Q}=0$,
whose solutions are given by  harmonic functions. 
Given another commuting isometry, each of such 
$U(1)\times U(1)$-invariant hyper-K\"ahler metrics is described by 
a harmonic function depending upon two variables, like in ref.~\cite{ov} ---
see sect.~4.

The hyper-K\"ahler geometry arises in the limit when N=2 supergravity 
decouples, because {\it any} 4d NLSM with rigid N=2 supersymmetry has 
 a hyper-K\"ahler metric \cite{fal}. The existence of such approximation 
is dependent upon the validity of eq.~(3.10). Otherwise, the hyper-K\"ahler 
limit may not exist. Given an $U(1)\times U(1)$ isometry of the hyper-K\"ahler
 metric under consideration, the existence of a translational (or 
tri-holomorphic) isometry does not pose a problem, since there
always exists a linear combination of two commuting abelian isometries 
 which is tri-holomorphic \cite{gib}. Some explicit examples of the 
correspondence between four-dimensional hyper-K\"ahler and quaternionic 
metrics were derived, e.g., in refs.~\cite{iv,iva} from harmonic superspace. 

We are now in a position to rewrite the classical UH metric (2.20) into the 
Tod form (3.4) by using {\it the same} coordinates 
as in eq.~(2.20).  We find
$$ P=1~,\quad e^u=\r~,\quad {\rm and}\quad d\wedge \hat{\Q}
= d\n\wedge d\m~~,\eqno(3.12)$$
which are all agree with eqs.~(3.5a), (3.5b) and (3.5c), respectively. 
Eq.~(3.5a) also implies that
$$ \L=3~.\eqno(3.13)$$

The classical UH metric does {\it not} have a hyper-K\"ahler limit because 
$\pa_{\r}u=1/\r$ is not proportional to $P=1$, so that eq.~(3.10)
is not valid. We confirmed this fact by direct checking that 
$\r^2 ds^2_{\rm FS}$ is K\"ahler and scalar-flat, but it is {\it not} 
Ricci-flat, and, hence, it is {\it not} hyper-K\"ahler.  This is in agreement 
with refs.~\cite{iv,gal} where the same conclusion was derived by other means.
It may have been expected because the dilaton N=2 supermultiplet (UH) is known
 to be dual to the N=2 supergravity multiplet under the $c$-map \cite{one}. We
 can now identify the $\r$ and $t$ coordinates in eq.~(2.14) with the $\r$ and
 $t$ coordinates used here, as well as to write down
$$ C = \m +i\n~~.\eqno(3.14)$$
The classical UH story is now complete. The $U(1)$-symmetric non-perturbative 
hypermultiplet metrics (with instanton corrections) are governed by 
non-separable solutions to the $SU(\infty)$ Toda equation (3.5b) 
with $P\neq 1$ \cite{plb,pro}, however, they are very difficult to find.

Fortunately, in the UH case, there is the second (linearly independent) abelian
 isometry. Given two abelian isometries commuting with each other, as in 
eq.~(3.2), one can write down an Ansatz for the UH metric in adapted 
coordinates where both isometries are manifest (in terms of a potential 
depending upon {\it two} coordinates only), and then impose the Einstein-Weyl 
conditions (3.3). Surprisingly enough, this programm was successfully 
accomplished in the mathematical literature only recently by Calderbank and 
Petersen \cite{cp}.   

The main result of ref.~\cite{cp} is the theorem that {\it any} 
four-dimensional quaternionic metric (of non-vanishing scalar 
curvature) with two linearly independent Killing vectors can be 
written down in the from 
$$\eqalign{ 
ds^2_{\rm CP} ~=~ &  \fracmm{4\r^2(F^2_{\r}+F^2_{\h})-F^2}{4F^2}\,
\left(\fracmm{d\r^2+d\h^2}{\r^2}\right) \cr 
 & + \fracmm{ [(F-2\r F_{\r})\hat{\a}-2\r F_{\h}\hat{\b} ]^2 +[-2\r F_{\h}
\hat{\a}
+(F+2\r F_{\r})\hat{\b}]^2 }{F^2[4\r^2(F^2_{\r}+F^2_{\h})-F^2] }~,\cr}
\eqno(3.15)$$
in some local coordinates $(\r,\h,\q,\j)$ inside an open region of the 
half-space $\r>0$, where $\pa_{\q}$ and $\pa_{\j}$ are the two Killing 
vectors, the one-forms $\hat{\a}$ and $\hat{\b}$ are given by
$$ \hat{\a}= \sqrt{\r}\,d\q\quad {\rm and}\quad \hat{\b}=\fracmm{d\j 
+\h d\q}{\sqrt{\r}}~~,\eqno(3.16)$$
while the whole metric (3.15) is governed by the function 
(= {\it pre-potential}) $F(\r,\h)$ obeying a linear differential equation,
$$\D_{\ch}F \equiv \r^2\left(\pa^2_{\r}+\pa^2_{\h}\right)F =
\fracmm{3}{4}F~~.\eqno(3.17)$$

Some comments are in order. 

The remarkable thing about eq.~(3.17) is the non-trivial fact that it is just  
a consequence of 4d, local N=2 supersymmetry. It is also remarkable and highly 
non-trivial that the {\it linear} equation (3.17) governs the solutions to the
highly non-linear Einstein-Weyl system (3.3) having an $U(1)\times U(1)$ 
 isometry.

We verified by using Mathematica that the {\it Calderbank-Petersen} (CP) Ansatz
(3.15) obeys the Einstein-Weyl equations (3.3) under the conditions (3.16) and 
(3.17). Hence, the metric (3.15) is quaternionic indeed. Our calculation also 
showed that the metric (3.17) has a {\it negative} scalar curvature provided 
that  
$$ 4\r^2(F^2_{\r}+F^2_{\h}) > F^2>0~~.\eqno(3.18)$$

After a field redefinition \cite{iva},
$$ G= F\sqrt{\r}~, \eqno(3.19)$$
the CP Ansatz (3.15) takes the form
$$-ds^2= G^{-2}\left\{ \fracmm{1}{P}(d\j +\hat{\Q})^2 + Pd\g^2\right\}~,
\eqno(3.20)$$
where \cite{iva}
$$ P= 1- \fracmm{GG_{\r}}{\r(G^2_{\r}+G^2_{\h})}~,\quad \hat{\Q}=
\left(\fracmm{GG_{\h}}{G^2_{\r}+G^2_{\h}}-\h\right)d\q~,\eqno(3.21)$$
and
$$ d\g^2 \equiv \r^2 d\q^2 + (G^2_{\r}+G^2_{\h})(d\r^2+d\h^2)~~.\eqno(3.22)$$
The Ansatz (3.20) is similar to the Tod Ansatz (3.4), while it allows one to 
identify the $G$ function (3.19) with the Tod coordinate $\r$ in eq.~(3.4). 
Plugging eq.~(3.19) into eq.~(3.17) yields a linear differential equation on
 $G(\r,\h)$ \cite{iva},
$$ \left(\pa^2_{\r}+\pa^2_{\h}\right)G = \fracmm{1}{\r}\pa_{\r}G~~.
\eqno(3.23)$$
Unfortunately, this does not seem to imply any direct relation between the 
Toda potential $u$ in eq.~(3.4) and the pre-potential $F$ in 
eq.~(3.17), because yet another reparametrization is needed to put 
the Ansatz (3.20) into the Tod form  (3.4).

The linear equation (3.17) means that the pre-potential $F$ is a local 
eigenfunction (of the eigenvalue $3/4$) of the two-dimensional 
Laplace-Beltrami operator 
$$\D_{\ch} =\r^2(\pa^2_{\r}+\pa^2_{\h}) \eqno(3.24)$$
on the hyperbolic plane $\ch$ equipped with the metric 
$$ ds^2_{\ch}= \fracmm{1}{\r^2}( d\r^2 +d\h^2)~,\quad \r > 0~. 
\eqno(3.25)$$  
Unlike the non-linear Toda equation (3.5b), the linearity of eq.~(3.17) allows
 a superposition of any two solutions to form yet another solution.  
In physical terms, this means that D-instantons have the cluster 
decomposition. This feature is not apparent in curved spacetime.

Equation (3.17) also has some scaling symmetries,
$$ \r\to \l_1\r~,\quad \h\to \l_1\h~,\quad F\to \l_2 F~,\eqno(3.26)$$
where we have introduced two rigid real parameters $\l_1$ and $\l_2$. 

To find the pre-potential $F$ of the classical FS metric of the UH in 
eq.~(2.21), we put the equivalent metric (2.27) into the form (3.22) by the 
coordinate change 
$$ \tilde{s} = \fracmm{4}{\sqrt{\h^2+\r^2}}~~.\eqno(3.27)$$
The corresponding Tod function in eq.~(3.20) appears to be
$$ G(\h,\r)= \sqrt{\h^2+\r^2} +\fracmm{1}{4}\left( 
\sqrt{(\h-1)^2+\r^2}+\sqrt{(\h+1)^2+\r^2}\right)~,\eqno(3.28)$$
whereas the associated pre-potential $F$ now follows from eq.~(3.28),
$$ F_{\rm FS}(\r,\h) = \fracmm{1}{\sqrt{\r}} \left\{
\sqrt{\h^2+\r^2}+ \fracmm{1}{4}\sqrt{ (\h-1)^2+\r^2} +
 \fracmm{1}{4}\sqrt{(\h+1)^2+\r^2}\right\}~~,\eqno(3.29)$$
where we have rescaled the coordinates $(\r,\h)$ and the functions $G$ and $F$
by using eq.~(3.26), in order to bring our result into the simplest form.
It is easy to verify that the function (3.29) is a solution to eq.~(3.17).

\section{Ooguri-Vafa solution}

The Ooguri-Vafa solution \cite{ov} describes the D-instanton corrected moduli
space metric of a matter hypermultiplet in flat spacetime. Nevertheless, it is 
also relevant to the UH case in curved spacetime, as regards the physical 
interpretation of the coordinates, the general structure of D-instanton 
corrections, and the linearization procedure to the metric equations.

Let's determine the physical meaning of the coordinates $(\r,\h)$
in eqs.~(3.17) and (3.29), in the context of the type-IIA superstring 
compactification on a CY threefold. 
 When N=2 supergravity is switched off, a generic 
$U(1)\times U(1)$ quaternionic NLSM metric becomes hyper-K\"ahler, while one 
of its $U(1)$ isometries is tri-holomorphic \cite{book}. This hyper-K\"ahler 
metric can always be put into the Gibbons-Hawking form (3.11) whose 
pre-potential $P\equiv V(x,y,\h)$ with possible isolated singularities is 
governed by the equation 
$$ V^{-1}\D_3 V =0~,\eqno(4.1)$$
where we have introduced the Laplace operator $\D_3$ in the flat 3-dimensional 
space parametrized by the cartesian coordinates $(x,y,\h)$. It is worth 
noticing that eq.~(4.1) is a consequence of rigid N=2 supersymmetry and a
 tri-holomorphic $U(1)$ isometry of the metric.

Because of another $U(1)$ symmetry, when being rewritten in the cylindrical 
coordinates ($\r,\theta;\h)$, the pre-potential $V$ becomes independent upon 
$\theta$. Equation (4.1) was used by Ooguri and Vafa \cite{ov} in their 
analysis of the UH moduli space near a conifold singularity under the  
$U(1)\times U(1)$ symmetry. The conifold singularity arises in the limit of 
the vanishing period, 
$$ \int_{\cal C}\O\to 0~~,\eqno(4.2)$$
where the CY holomorphic 3-form $\O$ is integrated over a non-trivial 3-cycle 
${\cal C}$ of CY. The powerful singularity theory \cite{sin}  can then be 
 applied to study the universal behaviour of the hypermultiplet moduli space  
near the conifold limit. 

In the context of the UH and the type-IIA superstring CY compactification, the
coordinate $\r$ represents the `size' of the CY cycle ${\cal C}$ or, 
equivalently, the action of the D-instanton originating from the Euclidean 
D2-brane wrapped about the cycle ${\cal C}$ (see also sect.~1). The physical 
interpretation of the $\h$ coordinate is just the (absolute) expectation value
 of the RR-field or, equivalently, that of the integral of the RR three-form 
over ${\cal C}$. The cycle ${\cal C}$ can be replaced by a sphere $S^3$, since
 the D2-branes only probe the overall size of ${\cal C}$ \cite{ov}.

The pre-potential $V$ is {\it periodic} in the RR-coordinate $\h$ since 
the D-brane charges are quantized \cite{ov}. This periodicity should also be 
valid in curved spacetime, i.e. for the full pre-potential $F$ (sect.~5). We 
normalize the corresponding period to be $1$, as in ref.~\cite{ov}. The 
Euclidean D2-branes wrapped $m$ times around the sphere $S^3$ couple to the RR
 expectation value on $S^3$ and thus should produce the additive contributions
 to both pre-potentials, $V$ and $F$, with the factor of $\exp(2\p im\h)$ each.

In the {\it classical} hyper-K\"ahler limit, when both N=2 supergravity and 
the D-instanton contributions are suppressed, the pre-potential $V(\r,\h)$ of
a single matter hypermultiplet cannot depend upon $\h$ since there is no 
perturbative superstring state with a non-vanishing RR charge. Accordingly, 
the classical pre-potential $V(\r)$ can only be the Green function of the 
two-dimensional Laplace operator in eq.~(4.1), 
$$ V = -\fracmm{1}{2\p}\log\r~~,\eqno(4.3)$$
in agreement with ref.~\cite{ov} (see also ref.~\cite{greene}). 

Unlike its hyper-K\"ahler counterpart (4.3), the classical (no D-instantons) 
quaternionic potential (3.29) depends upon $\h$ since the N=2 supergravity 
knows about the RR-fields. There is no direct correspondence between a 
quaternionic potential $F$ and a hyper-K\"ahler potential $V$: in particular,  
 the asymptotical behaviour of the potential (3.29) at $\r\to +\infty$ is 
different ($\propto \sqrt{\r}$) from eq.~(4.3). Moreover, the quaternionic 
metric associated to the potential (3.29) does not have a hyper-K\"ahler limit
at all (sect.~3).

The calculation of ref.~\cite{ov} to determine the exact D-instanton 
contributions to the hyper-K\"ahler potential $V$ is based on the idea 
\cite{bbs} that the D-instantons should resolve the singularity of the 
classical hypermultiplet moduli space metric at $\r=0$. A similar situation 
arises in the Seiberg-Witten theory of a quantized N=2 vector multiplet 
(see, e.g., ref.~\cite{book} for a review).

Equation (4.1) can be formally thought of as the equation on the 
electrostatic potential $V$ of electric charges of unit charge in the 
Euclidean upper half-plane $(\r,\h)$, $\r>0$, which are distributed along the 
axis $\r=0$ in each point $\h=n\in {\bf Z}$, while there are no two charges at
 the same point \cite{ov}. A solution to eq.~(4.1) obeying these conditions is 
unique, 
$$ V\low{\rm OV}(\r,\h)= \fracmm{1}{4\p} \sum^{+\infty}_{n=-\infty}\left(
\fracmm{1}{\sqrt{\r^2+ (\h-n)^2}}-\fracmm{1}{\abs{n}}\right)+{\rm const.}
\eqno(4.4)$$
After the Poisson resummation eq.~(4.4) takes the form \cite{ov}
$$ V\low{\rm OV}(\r,\h)=\fracmm{1}{4\p} \log\left( \fracmm{\m^2}{\r^2}\right)+
\sum_{m\neq 0}\fracmm{1}{2\p}e^{2\p im\h}\,K_0\left(2\p \abs{m}\r\right)~,
\eqno(4.5)$$
where the modified Bessel function $K_0$ of the 3rd kind has been introduced,
$$ K_s(z)=\fracm{1}{2}\int^{+\infty}_0\fracmm{dt}{t^{s-1}}\exp\left[
-\,\fracmm{z}{2}\left( t+\fracmm{1}{t}\right)\right]~,\eqno(4.6)$$
for all Re\,$z>0$ and  Re\,$s>0$, while $\m$ is a constant.

The asymptotical expansion of the Bessel function $K_0$ near $\r=\infty$ 
 now yields 
\cite{ov}
$$\eqalign{
 V\low{\rm OV}(\r,\h)~=~&\fracmm{1}{4\p} \log 
\left( \fracmm{\m^2}{\r^2}\right) +
\sum_{m=1}^{\infty} \exp \left(-2\p m\r\right)
\cos(2\p m\h)\times\cr
~& \times \sum_{n=0}^{\infty}\fracmm{\G(n+\fracm{1}{2})}{\sqrt{\p}n!
\G(-n+\fracm{1}{2})}\left(\fracmm{1}{4\p m\r}
\right)^{n+\frac{1}{2}}~~~.\cr}\eqno(4.7)$$

A dependence upon the string coupling constant $g_{\rm string}$ is easily
reintroduced into eq.~(4.7) after a substitution $\r\to\r/g_{\rm string}$.
The explicit factors of $\exp{(-m\r/g_{\rm string})}$ in eq.~(4.7) are the
expected D-instanton contributions in terms of the D-instanton action $(\r)$ 
\cite{ov}. 

\section{D-instantons in N=2 supergravity}

Our strategy for computing the D-instanton contributions to the UH moduli
space metric consists of two steps. First, we determine the {\it `basic'} 
solutions to the linear equation (3.17). Second, we sum the basic solutions 
after the action of the four-dimensional duality group $SL(2,{\bf Z})$. 
The $U$-duality group $SL(3,{\bf Z})$ is not expected to be a symmetry of our
solution since we ignore the five-brane instanton contributions.

The metric (3.25) in the hyperbolic plane $\ch$ is invariant under the action 
of the isometry group $SL(2,{\bf R})$ isomorphic to $SU(1,1)$. The hyperbolic 
plane $\ch$ can also be viewed as an open unit disc. In the electrostatic 
analogy (sect.~4) the electric charges are distributed on the boundary of this
 disc with unit density. 

A solution describing the infinite D-instanton sum in N=2 supergravity 
should be periodic in $\h$ (sect.~4), while we also expect it to be 
complete \cite{pro} --- by analogy with the Seiberg-Witten-type 
solutions. The quaternionic D-instanton sum should also respect the known 
limits: in the perturbative region (no D-instantons and no supergravity) 
it should reduce to eq.~(5.1) below, whereas near $\r=0$ it should have the UV 
fixed point (or the conformal infinity \cite{leb2}) where one expects the N=2 
superconformal field theory (or Landau-Ginzburg) description to be valid 
\cite{doug}. 

Though we cannot identify a dilaton in the full UH moduli space (the NLSM has 
general coordinate invariance in its target space), we can do it at weak 
coupling. Like instantons, a dilaton is the semiclassical notion.

The simplest solutions to eq.~(3.17) are given by power functions,
$$ P_s(\r,\h) =\r^s~.\eqno(5.1)$$ 
Substituting eq.~(5.1) into eq.~(3.17) yields a quadratic equation,
$$ s(s-1)=3/4~,\eqno(5.2)$$
whose only solutions are 
$$ s_1=3/2\quad {\rm and}\quad s_2=-1/2~.\eqno(5.3)$$

The special pre-potential (3.29) describes the classical (hyperbolic) metric 
via eq.~(3.15) with no D-instantons. A generic `multi-centre' solution to 
eq.~(3.17) with a {\it finite} instanton number $(m-2)$ was originally found 
by using the quaternionic-K\"ahler quotients of the $4(m-1)$-dimensional 
quaternionic projective space ${\bf HP}^{m-1}$ by an $(m-2)$-torus (i.e. by an
 $(m-2)$-dimensional family of commuting Killing vectors) with $m>1$ 
\cite{bf,gl,bgnp}. In our case the quaternionic projective plane 
${\bf HP}^{m-1}$ should be replaced by the non-compact quaternionic  
hyperboloid, ${\bf H}^{m-1}\to {\bf H}^{p-1,q}$, with $(p,q)=(m-1,1)$.

A generic multi-instanton solution to eq.~(3.17) with a torus isometry 
$U(1)\times U(1)$ is given by \cite{cp}
$$ F_m(\r,\h) = \sum_{k=1}^{m} \sqrt{ 
\fracmm{ a^2_k\r^2 +(a_k\h -b_k)^2}{\r}} \eqno(5.4)$$
with some real moduli $(a_k,b_k)$. Since the superposition principle applies, 
it is easy to check that eq.~(5.4) is a solution to eq.~(3.17) indeed. When 
the hyperbolic plane $\ch$ is mapped onto an open disc $D$, the `positions' of
 instantons are given by the marked points on the boundary of this disc where 
the torus action has its fixed points. The `twistors' (in the terminology of 
ref.~\cite{cp}) $\{a_k,b_k\}^{m}_{k=1}$ form the $2m$-dimensional vector space
 where the three-dimensional $SL(2,{\bf R})$ duality group naturally acts. 
In addition, the solutions (5.4) are merely defined modulo an overall real 
factor, while they obey eq.~(3.18) when $a_k\neq 0$. The total (real) 
dimension of the D-instanton moduli space $\cm_{m}$ (of instanton number 
$m-2$) is thus given by \cite{cp} 
$$ {\rm dim}\, \cm_m = 2m-{\rm dim}\,SL(2,{\bf R})-1 = 2m-4~.
\eqno(5.5)$$ 
The whole instanton moduli space (for all $m$) is clearly infinite dimensional,
 in agreement with the LeBrun theorem \cite{leb3}.

The pre-potentials $F$ of the four-dimensional quaternionic metrics of $m=2$ 
and $m=3$ were investigated e.g., in ref.~\cite{cp}. These metrics include, in
 particular, the quaternionic-K\"ahler extension of a generic four-dimensional 
hyper-K\"ahler metric with two centers and the $U(1)\times U(1)$ isometry 
(like the Taub-NUT and Eguchi-Hanson) \cite{iva}. In the case of $m=2$ one
always gets the hyperbolic metrics (no instantons). A {\it generic} case of 
$m=3$ can be put into the form \cite{cp}
$$ F_3(\r,\h)=\fracmm{1}{\sqrt{\r}}+\fracmm{(b+c/q)\sqrt{\r^2
+(\h+q)^2}}{\sqrt{\r}}+
\fracmm{(b-c/q)\sqrt{\r^2+(\h-q)^2}}{\sqrt{\r}}~~,\eqno(5.6)$$
where $(b,c)$ are two real (non-negative) moduli and $q^2=\pm 1$. Similar 
metrics with higher instanton numbers were investigated in 
refs.~\cite{gal,gl,ped,ag}. There always exist a UV fixed point (conformal
 infinity), and there are no unremovable singularities. All non-compact 
4-manifolds $X$ equipped with such metrics have negative first Chern class, 
while they naturally arise as resolutions of complex orbifold singularities 
\cite{csin}. At some values of the moduli a generic multi-instanton metric 
reduces to the classical metric governed by eqs.~(5.1) and (5.3). The  
holographic correspondence on the conformal boundary of $X$ is discussed 
in ref.~\cite{ket4}. 

The FS pre-potential (3.29) is special since it has no limit where N=2
supergravity decouples. In other words, eq.~(3.29) does not belong to the 
family (5.6), it does not reduce to any of the solutions (5.1) and (5.3) at 
weak coupling $\r\to\infty$, but it has the enhanced symmetry $SU(2,1)$ 
(sect.~2) that leaves no moduli and, hence, no instantons. 

The correspondence between eigenfunctions of the 2-dimensional Laplacian 
(3.24) in the hyperbolic plane $\ch$ and homogeneous solutions to eq.~(4.1) in
flat three dimensions is known to mathematicians \cite{csin}. A `basic' 
solution to eq.~(4.1) is given by Green's function in three dimensions,
$$\fracmm{1}{\sqrt{\r^2+\h^2}}~=~ \fracmm{1}{r}~.\eqno(5.7)$$
For example, the Ooguri-Vafa solution (4.4) is just the periodic 
(in $\h\to\h+1$) extension of the basic solution (5.7). The basic solution to 
eq.~(3.17), corresponding to the homogeneous solution (5.7), is 
given by \cite{csin}
$$ \sqrt{ \r +\fracmm{\h^2}{\r}}~~.\eqno(5.8)$$
In particular, the classical UH potential (3.29) is the sum of eq.~(5.8)
and the other two terms obtained from eq.~(5.8) by  $\h\to\h\pm 1$, up to a
normalization.  

To summarize this part of the section, we conclude that there are three basic
solutions to eq.~(3.17):
$$ \left\{  \fracmm{1}{\sqrt{\r}}~,\quad \sqrt{\r^3}~,\quad
 \sqrt{\r +\fracmm{\h^2}{\r}}\right\}~~.\eqno(5.9)$$

The action of $SL(2,{\bf Z})$ in the hyperbolic plane $\ch$ is given by
$$ \t ~\to~ \hat{\g}\t=\fracmm{a\t +b}{c\t +d}~,\quad (a,b,c,d) 
\in {\bf Z}~,\quad ad-bc=1~,\eqno(5.10)$$
where the complex coordinate 
$$\t = \t_1 + i\t_2\equiv \h + i\r\eqno(5.11)$$ 
has been introduced. In terms of $\t$ the basic solutions (5.9) take the form
$$ \fracmm{1}{\sqrt{\r}}=\fracmm{1}{\sqrt{{\rm Im}\,\t}} =
 \fracmm{1}{\sqrt{\t_2}}~,\eqno(5.12a)$$
$$\r^{3/2}=\left({\rm Im}\,\t\right)^{3/2}=(\t_2)^{3/2}~,\eqno(5.12b)$$
and
$$\sqrt{\r +\fracmm{\h^2}{\r}}=\sqrt{\fracmm{\t\bar{\t}}{{\rm Im}\,\t}} =
\fracmm{\abs{\t}}{\sqrt{\t_2}}~,\eqno(5.12c)$$
where $\abs{\t}^2=\t\bar{\t}=\t_1^2+\t^2_2=\r^2+\h^2$.

The most general $SL(2,{\bf Z})$-invariant solution to eq.~(3.17) is obtained 
by applying the discrete duality transformations (5.10) to all basic solutions 
(5.12) and summing over all these transformations (modulo a stability 
subgroup of each basic solution). 

As regards the power solutions (5.12a) and (5.12b), summing them over the
$SL(2,{\bf Z})$ gives rise to the non-holomorphic modular (automorphic) forms 
called the Eisenstein series \cite{ter}, 
$$E_s(\t,\bar{\t})=\sum_{\hat{\g}\in \Tilde{SL(2,{\bf Z})}}
P_s(\hat{\g}\t,\Bar{\hat{\g}\t})~,\eqno(5.13)$$
with $s=-1/2$ and $s=3/2$, respectively. To be well defined, the sum in 
eq.~(5.13) has to be limited to the quotient 
$$\Tilde{SL(2,{\bf Z})}=SL(2,{\bf Z}/\G_{\infty} \eqno(5.14)$$ 
with the stabilizer 
$$  \G_{\infty} =\left\{ \left(\begin{array}{cc} \pm 1 & * \\
0 & \pm 1 \end{array} \right)\in SL(2,{\bf Z})\right\}~.\eqno(5.15)$$
It is clear that the transformations (5.15) have to be excluded from the sum
in eq.~(5.13) since they leave the basic functions (5.12a) and (5.12b) 
invariant.

In the case of ${\rm Re}\,s>1$, the Eisenstein series (5.13) is given by 
\cite{ter}
$$ E_{s}(\t,\bar{\t})
=\fracmm{1}{2}\t_2^s\sum_{(p,n)=1}\fracmm{1}{\abs{p+n\t}^{2s}}=
\t_2^s +\t_2^s\sum_{(p,n)=1\atop n\geq 1}\fracmm{1}{\abs{p+n\t}^{2s}}~~,
\eqno(5.16)$$
where $(p,n)$ is the greatest common divisor of $p$ and $n$. The infinite sum
(5.16) can be interpreted as the contributions from the D-instantons of
discrete energy $p$ and discrete charge $n$ (see below). The duality-invariant 
sum of the basic solutions (5.12b) is thus given by the Eisenstein series 
(5.16) of $s=3/2$.

The Eisenstein series (as the functions of $s$) satisfy the functional equation
\cite{ter}
$$ \L(s)E_s(\t,\bar{\t})=\L(1-s)E_{1-s}(\t,\bar{\t})~,\eqno(5.17)$$
where
$$\L(s) =\p^{-s}\G(s)\z(2s)~.\eqno(5.18)$$ 
In particular, in the case of $s=3/2$, eqs.~(5.17) and (5.18) imply
$$ E_{-1/2}(\t,\bar{\t}) = \fracmm{3\z(3)}{\p^2} E_{3/2}(\t,\bar{\t})~,
\eqno(5.19)$$
where we have used the identities
$$ \G(\frac{3}{2})=\frac{1}{2}\p^{1/2}~,\quad
\G(-\frac{1}{2})=-2\p^{1/2}~,\quad \z(-1)=-\frac{1}{12}~~.\eqno(5.20)$$
Therefore, the duality-invariant sum over the basic solutions (5.12a) yields
{\it the same} result as that of eq.~(5.12b), namely, the Eisenstein series 
$E_{3/2}(\t,\bar{\t})$.

As regards the sum of (5.12c) over the $SL(2,{\bf Z})$, which is most important
for us, it also gives rise to {\it the same} function proportional to  
$E_{3/2}(\t,\bar{\t})$. This can be most easily seen by noticing that the
$SL(2,{\bf Z})$ transformations are generated by the T-duality transformation,
 $\h\to \h+1$, and the S-duality transformation,
$$ {\rm S}:\qquad \t ~\to~ -\fracmm{1}{\t}~~.\eqno(5.21)$$
When being applied to the basic solution (5.12a), the S-duality (5.21) yields
the basic solution (5.12c):
$$ \fracmm{1}{\sqrt{\r}}=\fracmm{1}{\sqrt{{\rm Im}\,\t}} 
~\stackrel{\rm S}{\longrightarrow}~ \fracmm{\abs{\t}}{\sqrt{{\rm Im}\,\t}}= 
\sqrt{\r +\fracmm{\h^2}{\r}}~~.\eqno(5.22)$$

A sum over $SL(2,{\bf Z})
$ always produces a periodic function of $\h$ of
period $1$. The Eisenstein series $E_s(\r,\h)$ has a Fourier series 
expansion \cite{ter}
$$\eqalign{
\L(s)E_s(\r,\h) ~=~& \r^s\L(s) + \r^{1-s}\L(1-s) \cr
~& +2\r^{1/2}\sum_{m\neq 0}\abs{m}^{s-1/2}\s_{1-2s}(m)K_{s-1/2}(2\p\abs{m}\r)
e^{2\p im\h}~~.\cr}\eqno(5.23)$$
Here $\s_s(m)$ is the so-called divisor function
$$ \s_s(m)=\sum_{0<d|m}d^s~,\eqno(5.24)$$
where the sum runs over all positive divisors $d$ of $m$. In the case of
$s=3/2$, eq.~(5.23) can be put into the form ({\it cf.} refs.~\cite{gg,gg2})
$$ 4\p E_{3/2}(\r,\h)= 2\z(3) \r^{3/2} +\fracmm{2\p^2}{3}
\r^{-1/2} + 8\p\r^{1/2}\sum_{m\neq 0 \atop n\geq 1} \abs{\fracmm{m}{n}}
e^{2\p imn\h}K_1(2\p\abs{mn}\r)~,\eqno(5.25)$$
where $\z(3)= \sum_{m>0}(1/m)^3$. The asymptotical expansion of the function
(5.25) for large $\r$ is given by ({\it cf.} eq.~(4.7))
$$\eqalign{
4\p E_{3/2}(\r,\h) = &  2\z(3)\r^{3/2} +\fracmm{2\p^2}{3}\r^{-1/2} 
+4\p^{3/2}\sum_{m,n\geq 1}\left(\fracmm{m}{n^3}\right)^{1/2}\left[
e^{2\p i mn(\h+i\r)}+e^{-2\p i mn(\h-i\r)}\right] \cr
  & \times \left[ 1 + \sum^{\infty}_{k=1}\fracmm{\G(k-1/2)}{\G(-k-1/2)}\,
\fracmm{1}{(4\p mn\r)^k}\right]~. \cr}\eqno(5.26)$$

We conclude that the infinite $SL(2,{\bf Z})$-invariant D-instanton sum of the
basic solutions (5.12) is always proportional to the Eisenstein series 
$E_{3/2}(\t,\bar{\t})$. Hence, the potential $F$ of the UH metric with all the
$D$-instanton contributions included is proportional to $E_{3/2}(\t,\bar{\t})$
too. This was presented as a conjecture in our earlier preprint \cite{preps}.
Our quaternionic D-instanton sum in the form of the Eisenstein series is a 
consequence of local N=2 supersymmetry, a toric isometry $U(1)\times U(1)$ and
 the $SL(2,{\bf Z})$ duality, by construction.

It is worth noticing that the restricted sum of the basic solutions (5.12c) 
over the T-duality (periodicity), $\h\to \h+1$, like in eq.~(4.4), is not 
enough to produce an S-duality invariant function.~\footnote{The author in 
grateful to the referee for pointing out a mistake in the earlier version of 
this \newline ${~~~~~}$ paper.}  

The Eisenstein series of eq.~(5.25) is known to appear in the exact 
non-perturbative description of the $R^4$ couplings in the ten-dimensional 
type-IIB superstrings \cite{gg}. When being expanded in the form (5.26),  this 
amounts to the infinite sum of the tree level, one-loop, and D-instanton 
contributions, respectively. We can, therefore, conclude that the 
non-perturbative UH moduli space metric with the D-instanton contributions is 
completely determined by the exact $R^4$ couplings in the ten-dimensional 
type-IIB superstrings, similarly to the one-loop contribution \cite{cand}. 
The one-loop correction proportional to $\r^{-1/2}$ has merely 
global meaning to the UH metric because the former can be 
removed by a NLSM field redefinition \cite{one}. At weak coupling, 
$\r\to\infty$, {\it both} D-instantions {\it and} N=2 supergravity decouple, 
so that $F_{\rm perturbative}\sim \r^{3/2}$. The limit where only D-instantons
 decouple is more tricky. It appears when the positions, 
$\h=\pm 2,\pm 3,\ldots$, of D-instantons are sent to the infinity, while the
reference points $\h=0,\pm 1$ are kept.

The exact $R^4$ couplings in the ten-dimensional type-IIB superstrings are 
known to be dictated by {\it the same} equation (3.17) due to ten-dimensional 
N=2 supersymmetry \cite{sethi}. In ref.~\cite{boris}, N=2 supersymmetry in 
eight dimensions was used to derive eq.~(3.17) --- see also ref.~\cite{berk}. 
In our case, N=2 local supersymmetry in {\it four\/} spacetime 
dimensions (i.e. merely eight supercharges) is used.

The classical FS metric is quaternionic and K\"ahler (sect.~2), whereas the 
quaternionic metric describing D-instantons is not K\"ahler. This means that 
our results for the UH coupled to N=2 supergravity cannot be rewritten to the 
N=1 supergravity form without truncation, since N=1 local supersymmetry in 4d 
requires the NLSM metric to be K\"ahler.
\vglue.2in

\section{The D-instanton-induced scalar potential}

When the UH is electrically charged, its scalar potential becomes non-trivial 
(we do not consider any magnetic charges here). This happens because of 
{\it gauging} of an abelian isometry in the UH moduli space. Abelian gauging 
in the {\it classical} UH target space was discussed in great detail
in ref.~\cite{pot} -- see also refs.~\cite{mich,tv,dall}. Since the 
D-instantons are supposed to preserve the $U(1)\times U(1)$ isometry of the 
classical UH moduli space, it is quite natural to gauge a $U(1)$ part of it
 in the presence of the D-instanton quantum corrections, in order to generate 
a {\it non-perturbative} UH scalar potential. The fixed points (zeroes) of the
 UH scalar potential determine new vacua in type-II string theory. An N=2 
gravity multiplet has an abelian vector field (graviphoton) that can be used
for abelian gauging. Otherwise, an extra N=2 vector multiplet is needed. 
In 4d, N=2 supergravity it is the quaternionic NLSM metric and its Killing 
vector that fully determine the corresponding scalar potential. We begin with 
the simpler case of 4d, {\it rigid} N=2 supersymmetry \cite{town,sakai}. 

Any hyper-K\"ahler N=2 NLSM in 4d can be obtained from its counterpart in 6d 
by dimensional reduction. No scalar potential for hypermultiplets
is possible in 6d. Hence, a non-trivial scalar potential can only be generated
 via a Scherk-Schwarz-type mechanism of dimensional reduction 
with a non-trivial dependence upon extra spacetime coordinates, like in
 refs.~\cite{town,sakai}
$$ \pa_4 \f^a = K^a(\f)~,\quad {\rm and} \quad \pa_5 \f^a =0~,\eqno(6.1)$$
where $K^a(\f)$ is a Killing vector in the NLSM target space with a 
hyper-K\"ahler metric $g_{ab}(\f)$ parametrized by four real scalars 
$\f^a$ and $a=1,2,3,4$. The Scherk-Schwarz procedure is consistent with rigid 
N=2 supersymmetry if and only if the Killing vector $K^a(\f)$ represents a 
translational (or tri-holomorphic) isometry, while there is always one such 
isometry in the case of an $U(1)\times U(1)$ symmetric hyper-K\"ahler metric. 
 Upon dimensional reduction of eq.~(6.1) down to 4d, the 6d NLSM kinetic terms 
produce the scalar potential 
$$ V(\f) = \fracmm{1}{2}\,g_{ab}K^aK^b \equiv \fracmm{1}{2}K^2~. \eqno(6.2)$$

Given a four-dimensional hyper-K\"ahler metric with a triholomorphic isometry,
 we are in a position to use the Gibbons-Hawking Ansatz (3.11) 
where this isometry is manifest, with $K^a=(1,0,0,0)$ and $\f^a=(t,\m,\n,\r)$.
  Equation (6.2) now implies  
$$ V = \fracmm{g_{tt}}{2}=\fracmm{1}{2}P^{-1}~~,\eqno(6.3)$$
where $P(\vec{X})$ is a harmonic functon of $\vec{X}=(\m,\n,\r)$. The 
(Gibbons-Hawking) multi-centre hyper-K\"ahler metrics are described by 
the harmonic function \cite{gh}
$$  P(\vec{X}) =\sum^m_{k=1} \fracmm {1}{\abs{\vec{X}-\vec{X}_k}}~~,
\eqno(6.4)$$
where the moduli $\vec{X}_i$ denote locations of the centers. The 
corresponding scalar potential (6.3) is
non-negative, while its absolute minima occur precisely at the fixed points 
where the harmonic function (6.4) diverges. Since $V=0$ at
these points, N=2 supersymmetry remains unbroken in all of these vacua. 
It is worth mentioning that the vacua are independent upon the NLSM 
parametrization used (the fixed points are mapped into themselves  under 
the NLSM reparametrizations). 

In the case of the UH in the 4d, N=2 supergravity, we have to deal with a 
quaternionic NLSM metric having a gauged abelian isometry. Hence, the
Gibbons-Hawking Ansatz (3.11) is to be replaced by the Tod Ansatz (3.4) 
\cite{plb}. We should also take into account the presence of the abelian gauge
 N=2 vector multiplet whose complex scalar component $(a)$ enters the NLSM 
scalar potential too. As was demonstrated in refs.~\cite{tv,dall}, the scalar 
potential in N=2 supergravity is a natural generalization of the scalar 
potential (6.2), 
$$ V = \fracmm{1}{{\rm Im}[\t(a)]}\,\fracmm{1}{2}K^2 =  
\fracmm{1}{{\rm Im}[\t(a)]}\,\fracmm{P^{-1}}{2\r^2}~~,\eqno(6.5)$$
where $\t(a)$ is the function governing the kinetic terms of the N=2 vector 
multiplet. We have used eq.~(3.4) in the second equation (6.5). See also a
recent paper \cite{sbl} for more general results.

The standard way of deriving the scalar potential in the gauged N=2 
supergravity uses the local N=2 supersymmetry transformation laws of the
fermionic fields (gauginos, hyperinos and gravitini) \cite{cgp,mich}. 
The contribitions of gauginos and hyperinos are always positive, whereas
the contribution of gravitini is negative. The results of refs.~\cite{tv,dall}
 imply that the negative (gravitini) contribution to the scalar potential 
cancels against the positive contributions due to the matter fermions (gaugino
 and hyperino) in the gauged N=2 supergravity. This is not the case in the 
gauged N=1 supergravity theories \cite{n1}. It may, therefore, not be possible
 to rewrite an N=2 gauged supergravity theory into the N=1 locally 
supersymmetric form without truncations ({\it cf.} our remarks at the end of 
sect.~5).

Because of unitarity of the N=2 supergravity theory effectively describing 
the unitary CY-compactifed  theory of type-II superstrings, there should be no 
ghosts in the N=2 vector multipet sector too, so that we should have  
$${\rm Im}[\t(a)]>0~~.\eqno(6.6)$$   
Being interested in the vacua of the effective N=2 supergravity theory, which 
are determined by the minima of its scalar potential (6.5), 
we do not need to know the function $\t(a)$ explicitly -- eq.~(6.6) is enough. 

In the classical approximation to the UH metric, eq.~(3.12) tells us that 
$P=const.$ This immediately gives rise to the run-away behaviour of the 
potential (6.5) with its absolute minimum at $\r=\infty$, in agreement with 
refs.~\cite{pot,dall}. This run-away solution is, of course, physically 
unacceptable because it implies the infinite CY volume i.e. a 
decompactification, as well as the `infinite' string coupling. One may hope 
that the use of the full (non-perturbative) UH metric may improve the scalar 
potential behaviour, because the D-instanton corrections imply $P\neq 1$ 
(sect.~3).  Unfortunately, finding the exact potential in this case amounts 
to solving the (non-linear, partial differential) Toda equation 
(3.5b), since the $P$-function is governed by the Toda potential via 
eq.~(3.5a). Though the 3d Toda equation is known to be integrable, it is
notorously difficult to find its explicit (non-separable) solutions.

When N=2 supergravity is switched off (after gauging), we can take the 
solution (4.4) or (4.5) for the $P$-function in the hyper-K\"ahler limit, 
i.e.
$$ P(\r,\h)=V\low{\rm OV}(\r,\h)~.\eqno(6.7)$$
In the perturbative region (large $\r$) the asymptotic expansion of the Bessel
 function in eq.~(4.5) again yields the infinite D-instanton sum (4.7),
$$\eqalign{
 P(\r,\h)~=~&\fracmm{1}{4\p}\log \left( \fracmm{\m^2}{\r^2}\right) +
\sum_{m=1}^{\infty} \exp \left(-\,\fracmm{2\p\abs{m\r}}{g_{\rm string}}\right)
\cos(2\p m\h)\cr
~& \times \sum_{n=0}^{\infty}\fracmm{\G(n+\fracm{1}{2})}{\sqrt{\p}n!
\G(-n+\fracm{1}{2})}\left(\fracmm{g_{\rm string}}{4\p\abs{m\r}}
\right)^{n+\frac{1}{2}} ~~~~,\cr}\eqno(6.8)$$
where the string coupling dependence has been reintroduced.

We conclude that the non-perturbative vacua of our toy model for the 
electrically charged UH in the presence of D-instantons are given by poles of 
the $P$-function defined by eqs.~(3.4) and (3.5). In the hyper-K\"ahler limit,
 the vacua are given by the fixed points of the D-instanton function (6.7) or
(6.8).
\vglue.2in

\section*{Acknowledgements}

The author would like to thank the Caltech-USC Center of Theoretical Physics 
in Los Angeles, USA, and the Institute of Theoretical Physics in Hannover,
Germany, for hospitality extended to him during the work on some parts of this
paper. Useful discussions with Stefan Vandoren and Pierre Vanhove and 
greatfully acknowledged. The author is also grateful to the referee for 
pointing out a mistake in the earlier version of this paper.

\end{document}